\documentclass{iopart} 
\usepackage{graphicx} 
 
\begin{document} 
 
\title{Multilayers of Zinc-Blende Half-Metals with Semiconductors}

\author{Ph Mavropoulos\dag 
\footnote[3]{To whom correspondence should be addressed, e-mail: 
Ph.Mavropoulos@fz-juelich.de}, I Galanakis\ddag\ and P H 
Dederichs\dag} 
 
\address{\dag\ Institut f\"ur Festk\"orperforschung, 
Forschungszentrum J\"ulich, D-52425 J\"ulich, Germany} 
\address{\ddag\ Institute of Microelectronics, NCSR ``Demokritos'', 
15310 Aghia Paraskevi, Athens, Greece} 
 
\begin{abstract} 
  We report on first-principles calculations of multilayers of 
  zinc-blende half-metallic ferromagnets CrAs and CrSb with III-V and II-VI 
  semiconductors, in the [001] orientation. We examine the ideal and 
  tetragonalised structures, as well as the case of an intermixed 
  interface. We find that, as a rule, half-metallicity can be 
  conserved throughout the heterostructures, provided that the 
  character of the local coordination and bonding is not disturbed. At 
  the interfaces with semiconductors, we describe a mechanism that can 
  give also a non-integer spin moment per interface transition atom, 
  and derive a simple rule to evaluate it. 
\end{abstract} 
 
 
\maketitle 
 
\section{Introduction} 
 
Half-metallicity is the property of some spin polarised materials to
exhibit a metallic density of states for one spin direction but a
clear band gap around the Fermi level $E_F$ for the
other~\cite{deGroot83}. Well known half metals are some Heusler
alloys~\cite{deGroot83,Galanakis02a,Galanakis02b,Galanakis04}, the
compound CrO$_2$~\cite{Korotin97}, the manganite
La$_{0.7}$Sr$_{0.3}$MnO$_3$~\cite{Soulen98}, magnetite (Fe$_3$O$_4$)
\cite{Yanase84} or some diluted magnetic semiconductors (DMS)
\cite{Akai98}. In recent years technological interest in possible
applications on spin-dependent transport has fueled the research in
this field.  For instance, ordered zinc-blende CrAs and CrSb have been
fabricated by molecular beam epitaxy, and experiment and calculations
suggest half-metallicity \cite{Akinaga00,Mizuguchi02b}. The high Curie
temperature $T_C$ (over 400~K in experiment \cite{Zhao01,Mizuguchi02}
and in theory \cite{Kubler03}) makes these compounds attractive also
for applications.  CrAs/GaAs \cite{Mizuguchi02b} and CrSb/GaAs
\cite{Zhao04} multilayers have been also realised, showing that
coherent heterostructures are possible.
 
Such zinc-blende compounds of transition elements with group-V and VI 
atoms (called pnictides and chalcogenides) have been also studied by 
{\it ab-initio} methods. Calculations of the total energy and 
structural optimisation agree with the experimental result that the 
ground state structure of such compounds is not the zinc-blende, but 
in most cases the NiAs structure \cite{Landolt-MY}---rather more complicated and 
certainly not half-metallic 
\cite{Wei87,Sanvito00,Continenza01,Zhao02,Xie03a}. The zinc-blende 
structure can only be achieved by epitaxial growth on semiconductors 
for a few monolayers. Heterostructures of half-metals with 
semiconductors are technologically interesting, since in principle they 
can be used to achieve spin injection from the ferromagnetic electrode to  
the semiconductor with 100 \%\ spin polarised current, a very useful quality
for potential application in spin transistors \cite{Datta}. 

Until now, among the high-$T_C$ half-metals, heterostructures of
Heusler alloys with semiconductors have been found to suffer from
interface states within the half-metallic gap
\cite{deGroot01,Picozzi03}, which can strongly reduce the spin
polarisation of the current.  On the other hand, zinc-blende
half-metals follow the structure and the bonding of the zinc-blende
semiconductors in a more coherent way, and thus one can expect to
avoid spurious interface states.  Motivated by these ideas, we
investigate in this paper the half-metal/semiconductor (HM/SC) contact
for the zinc-blende materials and see under which conditions
half-metallicity is preserved; moreover, we investigate the magnetic
moments at the interface.

The paper is organised as follows: In section~\ref{Sec:Method} we 
give a few details on our calculations. In section~\ref{Sec:Bulk} 
we shortly review the bulk properties of these half-metallic compounds,  
and address the effect of lattice distortion by tetragonalisation. In 
section~\ref{Sec:Multilayers} we present calculations on HM/SC 
multilayers in ideal geometry and epitaxy and see in which cases 
half-metallicity is preserved. In section~\ref{Sec:Interface} we 
examine interdiffusion effects at the interface and in 
section~\ref{Sec:Moments} we discuss the spin magnetic moments at 
the interface. Finally, we conclude with a summary in 
section~\ref{Sec:Summary}. 
 
\section{Method of calculation \label{Sec:Method}} 
Our calculations are based on density-functional theory within the
local density approximation (LDA) for the exchange-correlation
potential, with the Vosko, Wilk and Nusair parameterisation
\cite{Vosko80}. We employ the full-potential screened
Korringa-Kohn-Rostoker (KKR) Green function method
\cite{Papanikolaou02}, where the correct shape of the Wigner-Seitz
cells is taken \cite{Stefanou90}. For the calculation of the charge
and spin density we integrate along a contour on the complex energy
plane, which extends from below the valence $s$ states of the $sp$
atom up to the Fermi level, using 42 energy points; the lower states
are treated as atomic core states. For the Brillouin zone (BZ)
integration we have used a $\textbf{k}$-space grid of
30$\times$30$\times$30 in the full BZ for the bulk calculations and a
$\mathbf{k}_\parallel$-space grid 20$\times$20 in the two-dimensional
full BZ for the interface calculations. We have used a cutoff of
$\ell_{\mathrm{max}}$=3 for the wavefuctions and Green functions. More
details are given in reference~\cite{Galanakis03}. The calculations
are performed within the scalar relativistic approximation, which accounts
for relativistic effects other than the spin-orbit coupling. The
effect of the latter on the half-metallic property is quite small for
CrAs and CrSb, as shown in reference~\cite{Mavropoulos04} (the
reduction of the spin polarisation at the Fermi level is of the order
of 0.5\% for CrAs and 1.5\% for CrSb).

Finally, throughout the paper we have used the experimental lattice 
parameters for the semiconductors (see references \cite{landolt1}, 
\cite{landolt2} and \cite{landolt3} for a review). 
 
\section{Bulk properties \label{Sec:Bulk}} 
 
\subsection{Ideal zinc-blende structure} 
 
The bulk properties of zinc-blende half-metals have been discussed 
in many papers in the past 
\cite{Sanvito00,Continenza01,Zhao02,Xie03a,Galanakis03,Shirai01,Galanakis02d,Zhang04,Fong04,Xie03b,Liu03,Pask03,Sanyal,Xie03c,Zhang03,Picozzi}. 
Here we will summarise what is already known, and use this as a 
step towards the understanding of the bulk properties under 
tetragonal distortion and of the interface properties. 
 
In the zinc-blende structure every atom has tetrahedral coordination, 
with its first neighbours being of the other atomic species. This 
symmetry splits the $d$ states of the transition-metal element (TM) in 
two subspaces: the $t_{2g}$ with 3-fold degeneracy ($d_{xy}$, 
$d_{yz}$, and $d_{xz}$ orbitals) and the $e_g$ with 2-fold degeneracy 
($d_{z^2}$ and $d_{x^2-y^2}$).  The $t_{2g}$ states of the TM 
hybridise with the $p$ states of the neighbouring $sp$ atom (group V or 
VI), forming hybrids of bonding (lower in energy) and antibonding 
(higher) nature. This hybridisation and bonding-antibonding splitting 
is crucial for the formation of the gap in these compounds, and it is 
a characteristic of the tetrahedral coordination --- in the hexagonal 
coordination of the energetically more stable NiAs structure, for 
instance, it is not present. The effect is already known as ``$p$-$d$ 
repulsion'' in TM impurities in zinc-blende semiconductors 
\cite{Zunger}. 
 
The hybrids form wide bands, which retain the bonding-antibonding 
character and are therefore accordingly separated in energy. On 
the other hand, the $e_g$ orbitals are practically non-bonding and 
form narrow bands, which remain energetically between the bonding 
and antibonding $t_{2g}$ states, separated from both. Meanwhile, 
the $s$ states of the TM are still higher in energy, pretty much 
like the $s$ state of the Ga atom in GaAs stays above $E_F$. Only 
in cases of heavy $sp$ anions, which have reduced ionicity, do 
these $s$ states come closer to $E_F$ but still stay above it for 
minority spin---see for instance a recent study on MnBi 
\cite{Xu02}. The $s$ states of the $sp$ anion, on the other hand, 
are anyhow very low, lower than the bonding $p$-$d$ hybrids. 
 
This energetical separation already creates a band gap, and is
assisted strongly by the exchange splitting due to spin magnetism.
Thus, after the low-lying bonding bands are filled, the $e_g$ bands of
only majority spin are progressively filled, while the corresponding
minority bands are pushed higher above $E_F$ by the exchange
splitting. As we change the valence of the TM atom toward the right in
the periodic table, more majority bands are filled; the magnetic
moment increases; thus the exchange splitting also increases keeping
the minority $e_g$ states high in energy. In this way $E_F$ remains
between the well-separated bonding bands and the $e_g$ bands, within a
band gap. As a typical example we present the band structure of CrSb
in figure~\ref{fig:1} (left) for energies around the gap.

The situation described above changes when the lattice is 
compressed. Then the extended majority spin states of $p$ character 
feel the volume reduction and push $E_F$ higher, so that in the end $E_F$ 
enters the minority spin conduction band. After this point, although 
the gap still exists, it is below $E_F$ and half-metallicity is 
lost. Therefore it is necessary to identify which compounds are 
half-metallic in their equilibrium lattice parameter, and of course on 
which semiconductors they might grow. In 
reference~\cite{Galanakis03} we have identified such systems, and we 
summarise the most important results here in table~\ref{table:1}.

\subsection{Tetragonal distortion of the zinc-blende structure} 
 
When a zinc-blende half-metallic compound is grown on a semiconductor
substrate (say on (001)), its lattice will be tetragonally distorted
so that it can assume the in-plane lattice constant of the SC while
approximately keeping its own atomic volume.  If the lattice match is
good enough, the distortion will be minimal, and the electronic
structure will change only slightly.  Here we shall investigate this
effect for the case of reasonable HM/SC lattice matching, in order to
show that half-metallicity is not destroyed by moderate
tetragonalisation.
 
\begin{figure} 
   \begin{center} 
  \includegraphics[width=12cm]{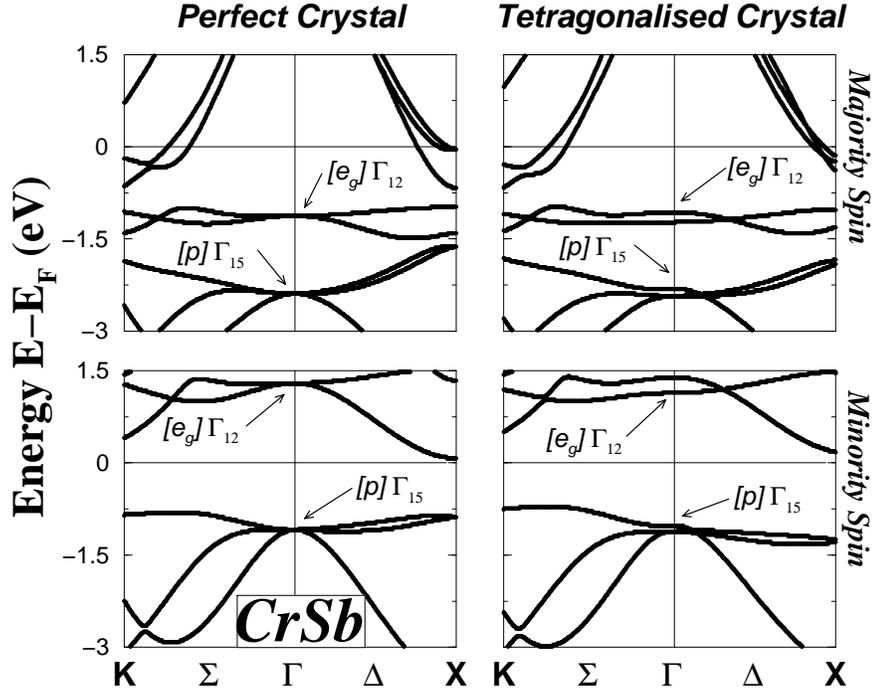} 
  \end{center} 
\caption{\label{fig:1}  
  Left: Band structure of zinc-blende CrSb in the equilibrium lattice
  parameter of 5.92~\AA .  Right: The same system, but with a moderate
  tetragonal distortion along the $z$ axis so as to obtain the ZnTe
  lattice constant in the (001) plane while keeping the equilibrium
  atomic volume. The triply degenerated $t_{2g}$ states at $\Gamma$
  has split up into two subspaces, one singly ($p_z$-$d_{xy}$) and one
  doubly degenerated ($p_z$-$d_{yz}$ and $p_y$-$d_{xz}$). Similarly,
  the doubly degenerated $e_g$ states ($d_{x^2-y^2}$ and $d_{z^2}$)
  have split into two subspaces. (Since the symmetry is no more cubic,
  the group representations at $\Gamma$ are no more $\Gamma_{12}$ and
  $\Gamma_{15}$, but we keep the notation for comparison with the
  cubic structure.) For moderate distortions, half-metallicity
  remains.}
\end{figure}

As a model system for the discussion we choose CrSb, which we assume
to be deposited on ZnTe (001) and to have taken the corresponding
in-plane lattice parameter of the semiconductor
a$_{\mathrm{ZnTe}}$=6.1 \AA\ and otherwise to have kept the atomic
volume.  This gives a ratio $c/$a=0.914, where $c$ is the lattice
parameter in the growth direction. The corresponding band structure is
shown in figure~\ref{fig:1} (right). Comparing to the CrSb bands in
the ideal zinc-blende geometry (figure~\ref{fig:1} (left)), we see
that the threefold degeneracy of the $t_{2g}$ states at $\Gamma$ has
split up into two subspaces, one singly ($p_z$-$d_{xy}$) and one
doubly degenerated ($p_x$-$d_{yz}$ and $p_y$-$d_{xz}$). Also the $e_g$
representation at $\Gamma$ has split in its constituents, $d_{z^2}$
and $d_{x^2-y^2}$. This behavior is expected because the $z$ axis is
now distinguished. Evidently half-metallicity is preserved by this
moderate distortion.
 
Such behavior is typical for all HM examined here. The
tetragonalisation can have an effect on half-metallicity if $E_F$ is
at the edge of the gap for the ideal structure. Then mainly the
extended $p$ states, feeling the lateral change, will push $E_F$
higher, similarly to the case of lattice compression; ultimately, if
the distortion is too large, half-metallicity can be destroyed. In
particular, if $c/a <1$ the $p_z$ states are squeezed and move higher
in energy, while if $c/a >1$ this happens to the $p_x$ and $p_y$
states.

\section{HM/SC multilayers \label{Sec:Multilayers}} 
 
In a recent examination of possible combinations of the transition 
elements V, Cr, and Mn, with group-V and VI elements 
\cite{Galanakis03}, we found that some of them are half-metallic at 
their equilibrium lattice constants, which fit also reasonably well to 
those of some semiconductors (SC). Here we examine the cases of 
CrAs/SC and CrSb/SC (001) multilayers for which the lattice mismatch 
should be small, as can be seen from table~\ref{table:1}. Note that 
within the LDA the lattice constant is usually underestimated by up to 
2-4\% \cite{Asato99}. Thus, if one allows for such a small increase of 
the calculated values of table~\ref{table:1} for the HM lattice 
constants, plus a moderate adjustment due to lattice mismatch, one 
arrives at the HM/SC combinations examined here. 
 
We assume that the structures have the experimental SC lattice 
constant and that the ideal zinc-blende structure is kept 
throughout. The systems consist of 4 monolayers (ML) of HM followed by 
4 ML of SC and periodically repeated, in accordance with the 
experimental result that only a few ML can exist within the 
periodically repeated supercell in the CrAs/GaAs case 
\cite{Mizuguchi02b}. In the $[$001$]$ direction of growth, this 
corresponds to interchanging monoatomic layers: e.g., for CrAs/GaAs, 
we have a supercell of the form ...Cr/As/Cr/As/Ga/As/Ga/As... . For 
test purposes we have performed calculations on 8ML HM/8ML SC, and 
seen that our conclusions remain unchanged. 
 
\begin{figure} 
   \begin{center} 
  \includegraphics[angle=270,width=16cm]{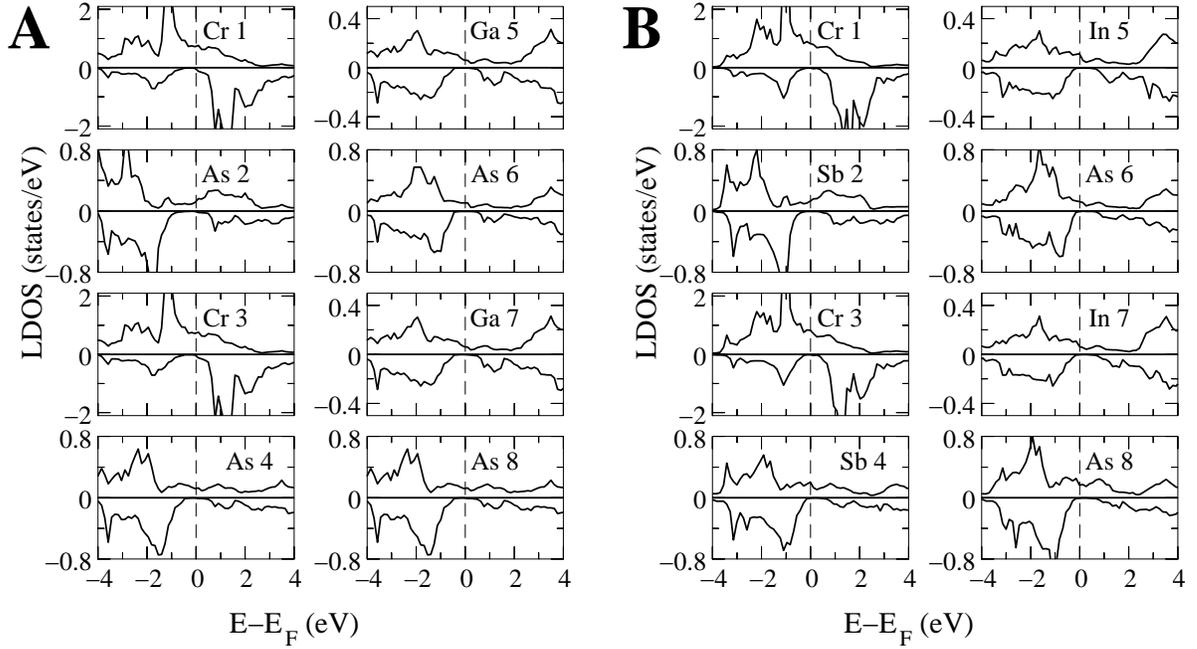} 
   \end{center} 
\caption{\label{fig:2}  
Atom-resolved DOS of CrAs/GaAs (A) and 
CrSb/InAs (B) (001) multilayers in supercell geometry. The inset 
numbers refer to the enumeration of successive layers. In CrAs/GaAs, 
the following atoms have equivalent environment and DOS: Cr1 and Cr3, 
As4 and As8, Ga5 and Ga7.} 
\end{figure}

Our results on the the local DOS (LDOS) are presented in
figure~\ref{fig:2} for CrAs/GaAs (left) and for CrSb/InAs (right). In
the multilayers we see that half-metallicity is conserved throughout.
This means that no interface states within the gap are formed for the
minority spin at the HM/SC interface. This can be understood since the
growth is coherent, so that the local environment of the interface Cr
atoms is not changed in CrAs/GaAs (This is also the case in CrAs/MnAs
multilayers; see, e.g., reference \cite{Fong04}). In CrSb/InAs each
interface Cr atom has two Sb neighbours on the one side and two As
neighbours on the other, but the minority gap remains, since the
$p$-$d$ hybridisation and the bonding-antibonding splitting are still
realised (the difference between having an As neighbour instead of an
Sb one is that the $p$ states of the former are somewhat lower in
energy).
 
For the majority spin the local DOS at $E_F$ decays within the SC 
layers. For a thick SC spacer it should vanish far from the interface, 
since the SC band gap is present and the DOS comes from exponentially 
decaying metal-induced gap states. 
 
The main difference between the CrAs/GaAs multilayer and the CrSb/InAs 
one is the position of $E_F$ within the minority gap. In CrAs/GaAs, 
$E_F$ just touches the conduction band, while for CrSb/InAs it is 
lower, closer to the middle of the gap.  This is a volume effect, 
clear also in the bulk CrAs case. The reduced lattice constant of 
CrAs/GaAs compared to CrSb/InAs affects the position of the $E_F$ 
within CrAs: the $p$ electrons feel the reduced volume and shift 
towards higher energy, pushing also $E_F$. 
 
\begin{figure} 
  \begin{center} 
  \includegraphics[angle=270,width=16cm]{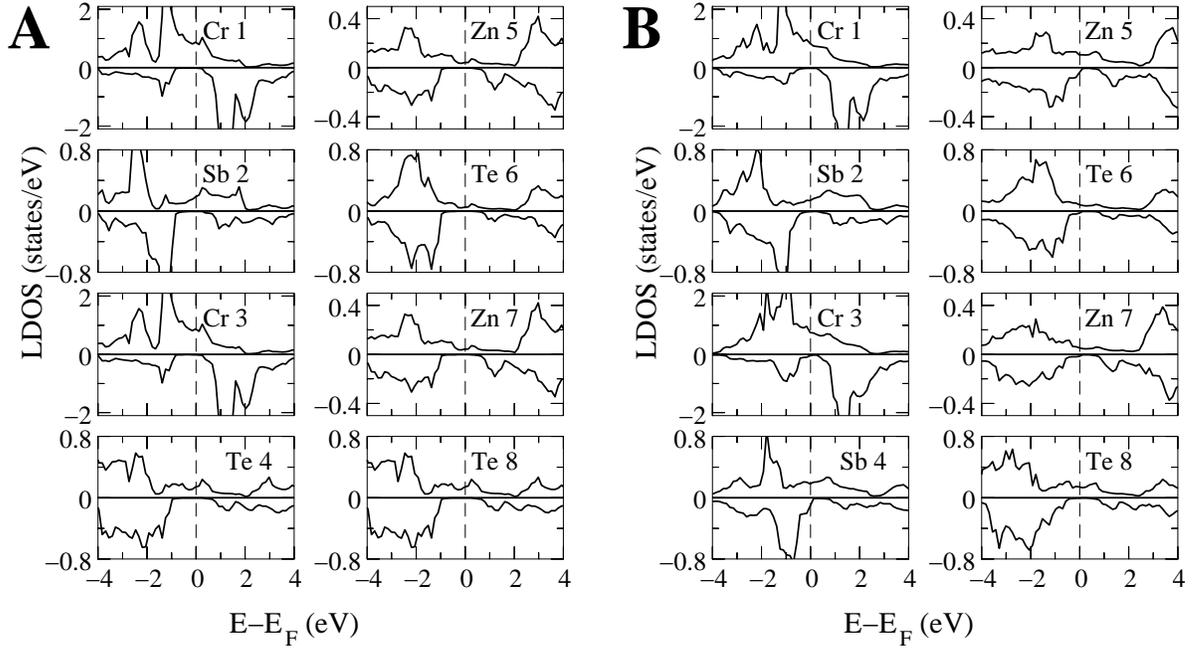} 
  \end{center} 
\caption{\label{fig:3}  
Atom-resolved DOS of CrSb/ZnTe (001) multilayer 
in supercell geometry, in two different configurations.  A: Multilayer 
of the form Cr/Sb/Cr/Te/Zn/Te/Zn/Te.  Half-metallicity is present 
throughout. B: Multilayer of the form Cr/Sb/Cr/Sb/Zn/Te/Zn/Te, where 
the difference from case A is that here a Sb/Zn contact is 
present. This destroys half-metallicity, as an interface state is 
formed at the Sb/Zn interface for the minority-spin states. This can 
be seen in particular at the Sb~(4) and Zn~(5) atom LDOS.} 
\end{figure}

Next we present the multilayer CrSb/ZnTe. We can call this a ``mixed 
valence'' multilayer, because it combines a 3d/group-V HM with a II/VI 
semiconductor. Thus, the environment of the interface Cr atom is 
highly anisotropic: on the one side it has two Sb neighbours and on 
the other two Te neighbours, the electronegativity of which is higher 
than Sb. Nevertheless, half-metallicity survives, since the $p$-$d$ 
repulsion creating the bonding-antibonding splitting is still there, 
(although with different strength between Cr-Sb and Cr-Te), and the 
strong exchange splitting is of course also present. The half-metallic 
LDOS is shown in figure~\ref{fig:3} (left). 
 
However, there are two inequivalent ways to construct the multilayer,
accounting for two different possible interfaces. The first is
Cr/Sb/Cr/Te/Zn/Te/Zn/Te, which is the half metallic one presented in
figure~\ref{fig:3} (left), and the second is Cr/Sb/Cr/Sb/Zn/Te/Zn/Te,
the difference being that here a Sb/Zn contact is present, i.e., a
III-VI hybrid. In the latter case half-metallicity is destroyed, due
to interface states which occur at the Sb/Zn interface at $E_F$ for
the minority spin LDOS. The corresponding LDOS is shown in
figure~\ref{fig:3} (right), where the minority-spin interface states
are formed at $E_F$ for the Sb~(4) and Te~(5) atoms.  Similarly, only
the Cr/Te/Cr/As/In/As/In/As stacking would retain half-metallicity in
the case of a CrTe/InAs system, since in the case of the
Cr/Te/Cr/Te/In/As/In/As stacking an In/Te interface is present.
 
Closing this section, we note that in the case of interfaces of
half-metallic Heusler alloys with SC, the half-metallic gap is usually
destroyed \cite{deGroot01,Picozzi03,Lezaic04}, e.g. from all the
(theoretically) studied cases in reference \cite{deGroot01} only the
NiMnSb(111)/CdS(111) interface retained the half-metallicity, and this
only when the interface was between the Sb and S atoms. Here, on the
contrary, the HM gap is, as a rule, present, since the nature of the
bonding is continued coherently at the interface.
 
\section{Intermixed interface \label{Sec:Interface}} 
 
Up to now we have considered only the ideal interface, without 
interdiffusion. However, in an experiment one can imagine that some 
intermixing can occur, and the question of keeping half-metallicity is 
raised again. As before, this can be discussed in terms of the bonding 
character. 
 
Consider, for instance, the CrAs/GaAs (001) interface, with an
intermixed CrGa monolayer (i.e., a layer sequence of
...Cr/As/Cr$_{0.5}$Ga$_{0.5}$/As/Ga/...), under the assumption that
the tetrahedral cation-anion coordination is not severely distorted
(this is a necessary requirement in all considerations of
half-metallicity in these materials). Then, the first neighbours of
each interface Cr atom will be As atoms, and the same holds for the
neighbours of the Ga atoms. A Cr-Ga bond will not be present, since
the closest distance between Cr and Ga atoms is the one of second
neighbours. Thus we expect the same bonding-antibonding scheme as
before, so that the HM gap should remain. Changes \emph{will} occur,
of course, for example due to the different electrostatic potential at
the interface, since Cr and Ga gave different ionicity. Thus $E_F$ can
be shifted with respect to the ideal interface.
 
\begin{figure} 
  \begin{center} 
 \includegraphics[width=8cm]{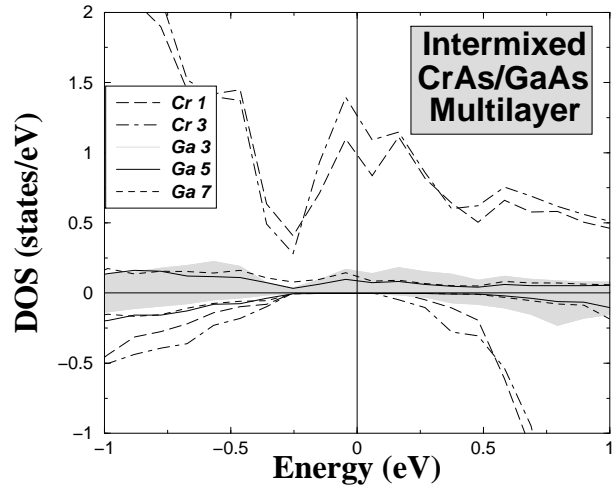} 
  \end{center} 
\caption{\label{fig:4} Atom-resolved DOS of CrAs/GaAs (001) 
multilayer in supercell geometry, in the case of an intermixed 
interface. Along the growth axis the successive layers are 
Cr/As/Cr$_{0.5}$Ga$_{0.5}$/As/Ga/As/Ga/As, thus in the third 
layer we consider both Ga and Cr sites (denoted as Cr 3 and Ga 3 
in figure labels). The system is half-metallic.} 
\end{figure}

In order to investigate if the HM property is really preserved, we 
performed a calculation of such an intermixed interface, by 
constructing an in-plane supercell, double in size than the usual 
cell and containing one Cr and one Ga atom at the interface. The 
DOS in the vicinity of $E_F$ is shown in figure~\ref{fig:4}. 
Evidently, the system is half-metallic.

\section{Magnetic moments \label{Sec:Moments}} 
 
In the bulk HM, the total spin moment per unit cell is integer (in 
$\mu_B$), as follows from just noticing that, due to the gap at 
$E_F$, the number of minority-spin electrons is an integer. This 
line of thinking leads to a simple ``rule of 8'' \cite{Galanakis03} 
connecting the total number of valence electrons 
$Z_{\mathrm{tot}}$ of the unit cell and the total spin moment 
$M_{\mathrm{tot}}$ per unit cell: 
\begin{equation} 
M_{\mathrm{tot}}= (Z_{\mathrm{tot}}-8)\ \mu_B, 
\label{eq:1} 
\end{equation} 
since one has a total of 8 electrons in the bonding $p$-$d$ bands and 
in the (deeper-lying) $s$ band of the $sp$ element, and the remaining 
the electrons occupy only majority states and build up the magnetic 
moment. This ``rule of 8'' is generally valid for the whole family of 
zinc-blende half-metallic compounds, expressing a Slater-Pauling 
behaviour; similar rules hold also for the Heusler 
alloys~\cite{Galanakis02a,Galanakis02b,Galanakis04} or HM 
superlattices~\cite{Fong04}. As regards to the distribution of the 
spin moment among the atoms, the $sp$ element has an induced moment 
opposite to the one of the transition element (thus for the local 
moment of the transition atom we have 
$M_{\mathrm{loc}}^{\mathrm{TM}}>M_{\mathrm{tot}}$). This can be 
understood in terms of the local character of the occupied 
bonding-$p$-$d$ majority and minority bands (majority are more 
$d$-like around the transition atom; minority are more $p$-like around 
the $sp$ atom). For details we refer to reference~\cite{Galanakis03}. 
Here we discuss the magnetic moments of the multilayers and 
interfaces. 
 
Since at the interface the translational symmetry is broken, the
magnetic moment per transition atom is no longer obliged to be an
integer, even if half-metallicity is preserved. This has been already
observed in the case of (001) surfaces \cite{Galanakis03}. In the case
of a transition-atom-terminated interface with a SC, the total moment
per interface transition atom depends on the valency of its neighbours
on each side. To understand this, it is convenient to consider the
occupied bonding $p$-$d$ hybrids plus the low-lying $s$ states of the
$sp$ atom as a kind of ``reservoir'' which has space for $6+2=8$
electrons (both spin directions are occupied), and to count the
electrons which fill it up; once the state reservoir is filled, the
remaining electrons will occupy only majority-spin states ($e_g$ and
$p$-$d$ antibonding) and build up the moment. Every $sp$ neighbour of
the TM introduces thus 8 low-lying states, and at the same time
$Z_{\mathrm{neighb}}$ valence electrons; this adds up to a total of
$8-Z_{\mathrm{neighb}}$ unfilled reservoir states, or holes, to absorb
the TM electrons. On the other hand, each one of these $sp$ neighbours
has a total of 4 neighbours itself, thus it contributes only $1/4$ of
its states to the particular TM. Adding these up, we find a modified
``rule of 8'', applicable for the total spin moment per interface or
surface TM in zinc-blende half-metals provided that half-metallicity
is preserved at the interface or the surface (eq.~(\ref{eq:1}) is a
special case of this):
\begin{equation} 
M_{\mathrm{interf/surf}} = \left( Z_{TM} + \sum_{\mathrm{neighb}=1}^N 
\frac{(Z_{\mathrm{neighb}}-8)}{4}   \right) \mu_B 
\label{eq:2} 
\end{equation} 
Here, $N$ is the number of neighbours, which will be less than 4 in
the case of a surface TM. As in the case of eq.~(\ref{eq:1}), this
simple rule applies only when half-metallicity is present. The moment
$M_{\mathrm{interf/surf}}$ found by this rule is not expected to be
completely localised at the TM, but rather distributed among the TM
and the surrounding sites.

\begin{figure} 
  \begin{center} 
  \includegraphics{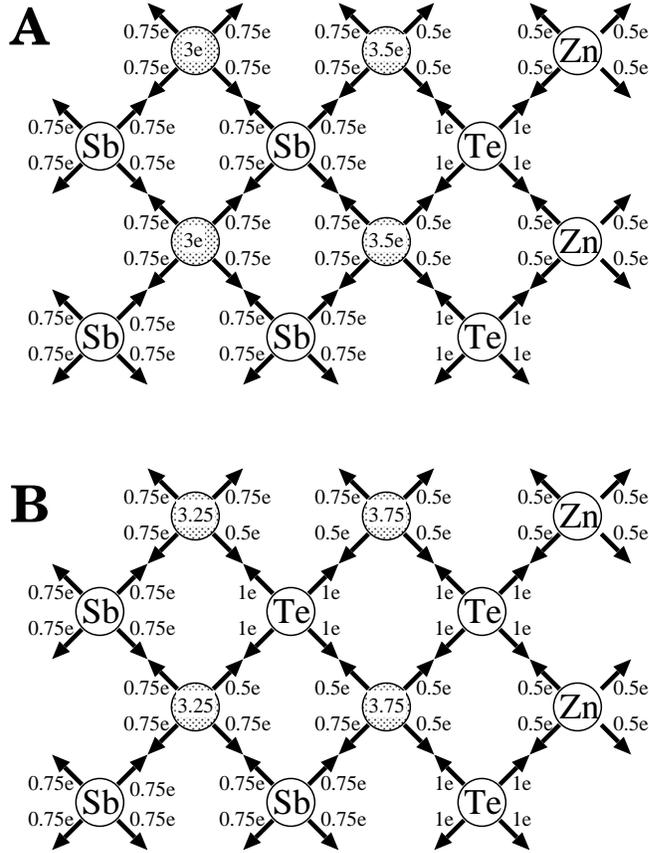} 
  \end{center} 
\caption{\label{fig:5} Schematic description for the non-integer 
  moment at the interface ZnTe/CrSb. Shaded circles stand for Cr
  atoms. Arrows indicate the electrons donated by each atom to the
  bonding $p$-$d$ bands. The remaining electrons at the Cr atoms
  occupy only majority spin states and build up the magnetic moment.
  In the bulk, one has 3~$\mu_B$ per Cr atom. A: Abrupt interface,
  where each interface Cr atom has two Sb and two Te neighbours. Since
  each Te atom donates an extra electron compared to the Sb atom, the
  interface Cr atom keeps an extra $1/2$~electron and increases its
  moment by 0.5~$\mu_B$. B: Intermixed interface. The interface Cr
  atom has one Sb and three Te neighbours, while the sub-interface Cr
  atom has three Sb and one Te neighbours. The interface and
  sub-interface Cr spin moments are increased by 0.75~$\mu_B$ and
  0.25~$\mu_B$, respectively, compared to the bulk CrSb value.}
\end{figure}

The situation is illustrated schematically in figure~\ref{fig:5}A for 
the case of ZnTe/CrSb (in the figure we do not discuss the $s$ states 
of the Sb and Te atoms, since they are energetically low-lying and 
already filled by two electrons). In the bulk, each Sb neighbour of 
the Cr atom gives 3/4 of an electron to the $p$-$d$ bond, so that the 
4 Sb neighbours give a total of 3 electrons per Cr atom. At the 
ZnTe/CrSb interface the Cr atom has two Sb and two Te neighbours.  The 
Sb ones give again 3/4 of an electron each, while the Te give one 
electron each (since they have one more valence electron).  Adding 
these up, we have a total of 3.5~$e$ donated by the $sp$ atoms to 
partly fill the bonding states of the interface Cr atom.  Since there 
are 6 such states, the remaining 2.5~$e$ are donated by the Cr atom, 
while the magnetic moment is built up by the $Z_{\mathrm{Cr}}-2.5=3.5$ 
remaining electrons of Cr, where $Z_{\mathrm{Cr}}=6$ is the valence of 
Cr. As a result, a half-integer moment of $M=3.5\ \mu_B$ per interface 
Cr atom occurs. In fact, this cannot be strictly localised at the 
interface, but is distributed among the interface atoms and their 
neighbours.  
 
Note that the above sum rules refer only to the total moments, but not
to the local ones. For instance, table~\ref{table:2} gives the local
spin moments for the half-metallic multilayers. In comparing CrAs/GaAS
(total interface moment: 3 $\mu_B$) with CrSb/ZnTe (total interface
moment: 3.5 $\mu_B$) results we find that the local interface Cr
moment increases by about 0.45 $\mu_B$ and the other 0.05 $\mu_B$ is
achieved by a reduction of the absolute value of the negative anion
moments; interface Te atoms have a spin moment of around -0.06 $\mu_B$
compared to the -0.10 $\mu_B$ of the As interface atoms. In the case
of the CrSb/InAs system the total interface moment should be 3 $\mu_B$
as for CrAs/GaAs and the larger Cr moment at the interface for the
former system is compensated by the larger absolute values of the spin
moments of the $sp$ anions.  The difference between the two systems
arises mainly due to the larger electronegativity of As compared to
Sb. The As $p$ states are originally lower than the Sb ones, so that
the $p$-$d$ hybridisation is weaker, and the bonding states of
majority spin are more localised around the As atom; thus the absolute
value of both Cr and As moments is smaller.

But it is also possible to obtain a value of $3.25\ \mu_B$ or $3.75\ 
\mu_B$ per Cr atom. Such an example is shown schematically in
figure~\ref{fig:5}B, where we have substituted the pure Sb layer at
the interface by a Sb$_{0.5}$Te$_{0.5}$ intermixed layer. Each Cr atom
at the interface has one Sb and three Te atoms as first neighbours,
and gives away 2.25 electrons to the bonding states. Thus
$Z_{\mathrm{Cr}}-2.25=3.75$ electrons are left to build up the spin
moment. Similarly the Cr atoms in the sub-interface layer have now
three Sb and one Te atoms as first neighbours and a total spin moment
of $3.25\ \mu_B$.
 
The above sum rule can be considered as a generalisation of the  
sum rule derived in reference \cite{Galanakis03} for the surfaces. In the case of  
the Cr-terminated CrAs(001) or CrSe(001) surfaces, the Cr atom 
at the interface loses two of its four neighbours and its spin moment  
is increased by 1.5 $\mu_B$ in the case of the CrAs(001) surface and by 
1.0 $\mu_B$ for the CrSe(001) surface. 
 
\section{Summary and conclusions \label{Sec:Summary}} 
 
We have performed first-principles calculations of (001) multilayers 
of half-metallic zinc-blende compounds (CrAs and CrSb) with III-V and 
II-VI semiconductors, focusing on the question whether 
half-metallicity is conserved at the interface. We have found that 
this can be the case under not too restrictive assumptions. 
 
Basically, the important requirement is that the coordination of the 
transition metal does not change at the interface: it should have four 
$sp$ neighbours of anionic (electronegative) character, as in the 
bulk. Then the bonding-antibonding splitting of the $p$-$d$ hybrides is 
retained and half-metallicity is conserved. It is not important that 
all four neighbours be of the same type. Thus, even in the InAs/CrSb 
and ZnTe/CrSb interfaces one observes half-metallicity. For the same 
reason, the intermixed Cr-Ga interface is half-metallic, too. We have 
also found that a moderate tetragonalisation does not affect the 
half-metallic character of the materials, thus growth of these 
half-metals on semiconductors with slightly different lattice constant 
brings no problems. 
 
We have also examined the magnetic moments at the interface. We have 
found that, in the case of an interface between a TM/group-V compound 
and a II-VI semiconductor, the total spin magnetic moment at the interface has a 
non-integer value. Based on simple arguments, applicable to 
any zinc-blende ferromagnetic surface or interface provided that 
half-metallicity is preserved, we have derived a sum rule (eq.~\ref{eq:2})  
to calculate this spin moment. 
 
We conclude that zinc-blende half-metallic compounds combine various 
qualities desirable for spintronics applications: half-metallic 
ferromagnetism, high Curie temperatures, coherent growth and 
half-metallic interfaces with semiconductors. We hope that our work is 
motivating for further experimental research in the field. 
 
\ack{Financial support from the RT Network ``Computational 
Magnetoelectronics'' (Contract RTN1-1999-00145) of the European 
Commission is greatfully acknowledged.}

\newpage 
 
\begin{table} 
\caption{ Zinc-blende compounds that are half-metallic (HM) at 
their equilibrium lattice parameter a (calculated within the LDA) 
together with semiconductors (SC) with lattice parameter close to 
that value (experimental values given). A $+$ means that the 
compound is HM at the lattice constant of the given SC, and a 
\fbox{$+$} means that the lattice mismatch is small, giving a 
candidate for epitaxial growth. For more details see 
reference~\cite{Galanakis03}.}\label{table:1} 
\begin{indented} 
 \item[] 
 \begin{tabular}{ccccccc} 
\br 
     & SC      & GaAs       & CdS  & CdSe       & InAs        & GaSb, ZnTe 
     \\ \br 
HM   & a(\AA ) & 5.65       & 5.82 & 6.05       & 6.06        & 6.10 \\ 
\mr 
VAs  & 5.65    & \fbox{$+$} & $+$  & $+$        &     $+$     & $+$  \\ 
VSb  & 5.98    & $-$        & $+$   & \fbox{$+$} &  \fbox{$+$} & \fbox{$+$} \\ 
CrAs & 5.52    & \fbox{$+$} & $+$  & $+$        &     $+$     & $+$  \\ 
CrSb & 5.92    & $-$        & $+$  & \fbox{$+$} &  \fbox{$+$} & \fbox{$+$} \\ 
\mr 
CrSe & 5.61    & $-$     &\fbox{$+$}& $+$       &     $+$     & $+$  \\ 
CrTe & 6.07    & $-$     &    $-$  & \fbox{$+$} &  \fbox{$+$} & \fbox{$+$}  \\ 
\br 
\end{tabular} 
\end{indented} 
\end{table} 
 
\begin{table} 
\caption{Local spin moments (in units of $\mu_B$ at the Cr and 
$sp$  atoms in the CrAs/GaAs, CrSb/InAs 
, and CrSb/ZnTe 
(half-metallic) 
  (001) multilayers.} \label{table:2} 
  \begin{indented} 
 \item[] 
\begin{tabular}{l|lr|lr|lr}\br 
Layer & 
\multicolumn{2}{c|}{CrAs/GaAs} & \multicolumn{2}{c|}{CrSb/InAs}  & 
\multicolumn{2}{c}{CrSb/ZnTe}\\ \mr 
1 & Cr: & 2.930 & Cr: & 3.195  & Cr: & 3.389\\ 
2 & As: &-0.200 & Sb: &-0.253  & Sb: & -0.230\\ 
3 & Cr: & 2.930 & Cr: & 3.095  & Cr: & 3.387\\ 
4 & As: &-0.107 & Sb: &-0.149  & Te: & -0.061\\ \mr 
5 & Ga: & 0.021 & In: &-0.005  & Zn: & 0.022\\ 
6 & As: &-0.011 & As: &-0.056  & Te: & -0.007\\ 
7 & Ga: & 0.021 & In: &-0.005  & Zn: & 0.022\\ 
8 & As: &-0.107 & As: &-0.194  & Te: & -0.061\\ \br 
\end{tabular} 
\end{indented} 
\end{table} 
 
\newpage 
 
\end{document}